\documentclass[]{spie}  %>>> use for US letter paper
\usepackage{amsmath,amsfonts,amssymb}
\usepackage{graphicx}
\usepackage[normalem]{ulem}
\usepackage[colorlinks=true, allcolors=blue]{hyperref}
\newboolean{english}
\newboolean{french}
\newboolean{german}
\newboolean{spanish}
\newboolean{italian}
\relax

\title{EMPOL: An EMCCD based Optical Imaging Polarimeter}

\author[a]{Shashikiran Ganesh}
\author[a,b]{Archita Rai} %<rai.archita@gmail.com>,
\author[a,b]{Aravind K} %<aravind139@gmail.com>
\author[a]{Alka Singh}
\author[a]{Prachi Vinod Prajapati}
\author[c]{Ashish Mishra} 
\author[a]{Prashanth Kasarla} %<prashanth@prl.res.in>,
\author[a]{Deekshya Roy Sarkar}
\author[a]{Pitamber Singh Patwal}
\author[a,b]{\\Namita Uppal}
\author[d]{Sunil Chandra}
\author[a]{Satyanarain Mathur}
\author[a]{Amish B Shah} %<abshah@prl.res.in>,
\author[a]{Kiran S Baliyan} %<baliyanprl@gmail.com>,
\author[a]{U C Joshi} %<ucjoshi1950@gmail.com>,
\affil[a]{Physical Research Laboratory, Navarangpura, Ahmedabad, India}
\affil[b]{Indian Institute of Technology Gandhinagar, Gandhinagar, India}
\affil[c]{University of Toledo, USA}
\affil[d]{Centre for Space Research, North-West University, Potchefstroom, South Africa}
\authorinfo{Further author information: (Send correspondence to S.G.)\\S.G.: E-mail: shashi@prl.res.in, Telephone: 91 79 26314508}

\begin{document} 
\maketitle

\begin{abstract}

An Andor 1K $\times$ 1K EMCCD detector has been used to develop an optical imaging polarimeter for use at the Cassegrain focus of 1.2~m telescope of PRL.  The optics is derived from an older single-element detector instrument and consists of a rotating half-wave plate as modulator and a Foster prism as an analyser.  The field of view of the instrument is 3 $\times$ 3 sq arcmin. We describe the instrument and the observational methodology in this document.  Extensive observations have been carried out with this instrument covering a large variety of sources e.g. near-Earth asteroids, comets, Lynds dark nebulae, open clusters and AGN such as blazars.  In the current communication, we discuss some results from the initial calibration runs while the other results will be presented elsewhere. 
 \end{abstract}

\keywords{EMCCD, Imaging polarimeter, optical polarisation, rapid modulation, astronomical polarimeter}

\section{INTRODUCTION}
\label{sec:intro}  % \label{} allows reference to this section

Polarimetry is a technique used in many areas of astronomy to study physical characteristics of a wide range of astronomical sources. In a number of researches, polarimetry plays a vital role in deriving interesting inferences which could not be revealed with any other technique. Understanding these advantages of the polarimetric techniques in astronomy, the Astronomy and Astrophysics Division of Physical Research Laboratory (PRL), India, has developed and been using an Optical Photo-polarimeter since the mid 1980s\cite{MRD1985BASI,UCJ1985MNRAS,UCJ1987AA,SG2009PRLTN}.  In the mid 1990s an effort was initiated to build an experimental instrument for infrared photo-polarimetry\cite{UCJ1997BASI,Manian1999BASI} using a single element InSb detector. The lack of a sensitive detector effectively limited the use of this infrared instrument to bright sources.  Later, with the availability of fast optical imagers with compact electronics, the opto-mechanical components were reused in a new instrument for optical imaging polarimetry by replacing the single element infrared detector Dewar with an EMCCD. This EMCCD based optical imaging polarimeter (EMPOL) has been in use since several years at the Mount Abu IR Observatory (MIRO). The polarization optics consists of a rotating half-wave plate (modulator) followed by a Foster prism (analyzer). The instrument has been used for various observations ranging such as some regions of the lunar surface, near-Earth asteroids, comets, open clusters, stars surrounding Lynds dark nebulae and active galactic nuclei such as blazars.

\noindent
Here we discuss the development of the instrument and present the details of the implementation with a few results from the characterisation observing runs.  Many observations have been completed with this instrument and the results will be presented elsewhere.  In the first section we list the  science cases of interest for observations with EMPOL.  In the next section the instrument development and implementation is described and the following section presents a few results from the calibration observations.  We conclude with a summary.

\section{Science cases for EMPOL linear polarimetric observations}

Study of many astronomical sources would benefit from an examination of their linear polarimetric properties.  Here we list a few of the science cases of interest for observations with EMPOL.

\subsection{Minor bodies of the solar system}

Comets and asteroids (including near-Earth asteroids) 
exhibit strong linearly polarised light arising from the scattering of sunlight by dust present in them.  Study of the polarisation properties as a function of wavelength helps to understand the characteristics of the dust grains.  

\subsection{Open clusters}
The light from the stars get partially polarized after passing through the dusty interstellar medium. The polarization provides the information about the line of sight dust distribution and the physical properties of the dust grains. The open clusters are spread out over the sky so the evolution of the physical parameters of the dust can be analysed over the region.  Linear Polarization is an excellent tool for determining the cluster membership. The procedure to identify memberships is based on the idea that the member stars are located behind common dust clouds which polarize the light, while most non-members maybe more randomly distributed in space and so their polarisation properties would be quite distinct. 

\subsection{Dark molecular clouds}

Dark clouds in the Milky Way are places where new stars may be forming in deeply embedded regions.  They are usually regions of dense molecular gas and dust.  Observations of the linear polarisation of starlight passing through these dense clouds are used to trace the magnetic fields in the region.  This can be used to trace the role of the magnetic fields in the star formation process.  

\subsection{Active Galactic Nuclei}

Active galactic nuclei have supermassive black holes at their centres.  These manifest their presence by emitting non-thermal highly polarised radiation.  The polarisation is also found to be variable and the timescale of variability provides an indication of the size of the emitting regions.  Oftentimes the energy emitted from an AGN emanates from a volume the size of the solar system and this energy is far larger than that from a whole normal galaxy.  Thus optical polarisation is a very important property to be monitored for AGN in order to understand their energy generation mechanisms.

\section{Instrument design and implementation}

\begin{figure} [hb]
   \begin{center}
\includegraphics[width=0.4\textwidth]{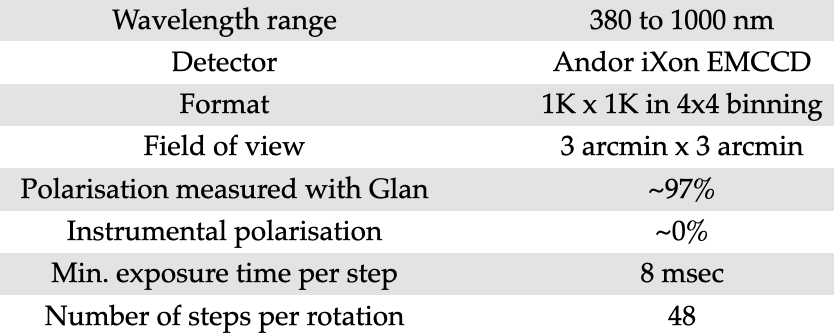}
    \end{center}
   \caption 
%>>>> use \label inside caption to get Fig. number with \ref{}
   { \label{fig:pars} 
Characteristics of EMPOL at a glance.}
\end{figure}

\noindent
The principle of the rapid modulation photo-polarimetric method\cite{Serkowski1974psns,FS1976ApOpt} is adopted in the current instrument.   Normally, in instruments with rapid modulation requirement, photomultiplier tubes(PMT) are used as detectors in photon-counting mode.  Rapid modulation allows to minimise or alleviate the atmospheric transmission variability affecting the polarisation measurement.  In the case of integrating detectors like conventional CCDs, such a fast modulation rate is not possible.  With the advent of electron multiplying CCD (EMCCD), near photon counting mode of detection has become available, with very high frame rates.  The characteristics of the EMPOL instrument, further discussed in the following sections, are shown in Fig. \ref{fig:pars}.

\subsection{Polarisation optics}
\begin{figure} [ht]
   \begin{center}
\includegraphics[scale=0.7]{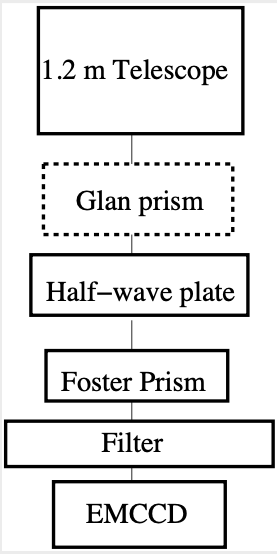}
    \end{center}
   \caption 
%>>>> use \label inside caption to get Fig. number with \ref{}
   { \label{fig:opticslayout} 
Schematics of the optical layout for EMPOL.}
\end{figure}

Imaging (linear) polarimetry requires array detectors with images taken through an analyser at 3 or more orientations.  
In our implementation of the linear imaging polarimeter, we employ a rotating half wave plate as modulator followed by a Foster prism as an analyser (see Fig. \ref{fig:opticslayout}).

\begin{figure} [ht]
   \begin{center}
\includegraphics[width=0.32\textwidth]{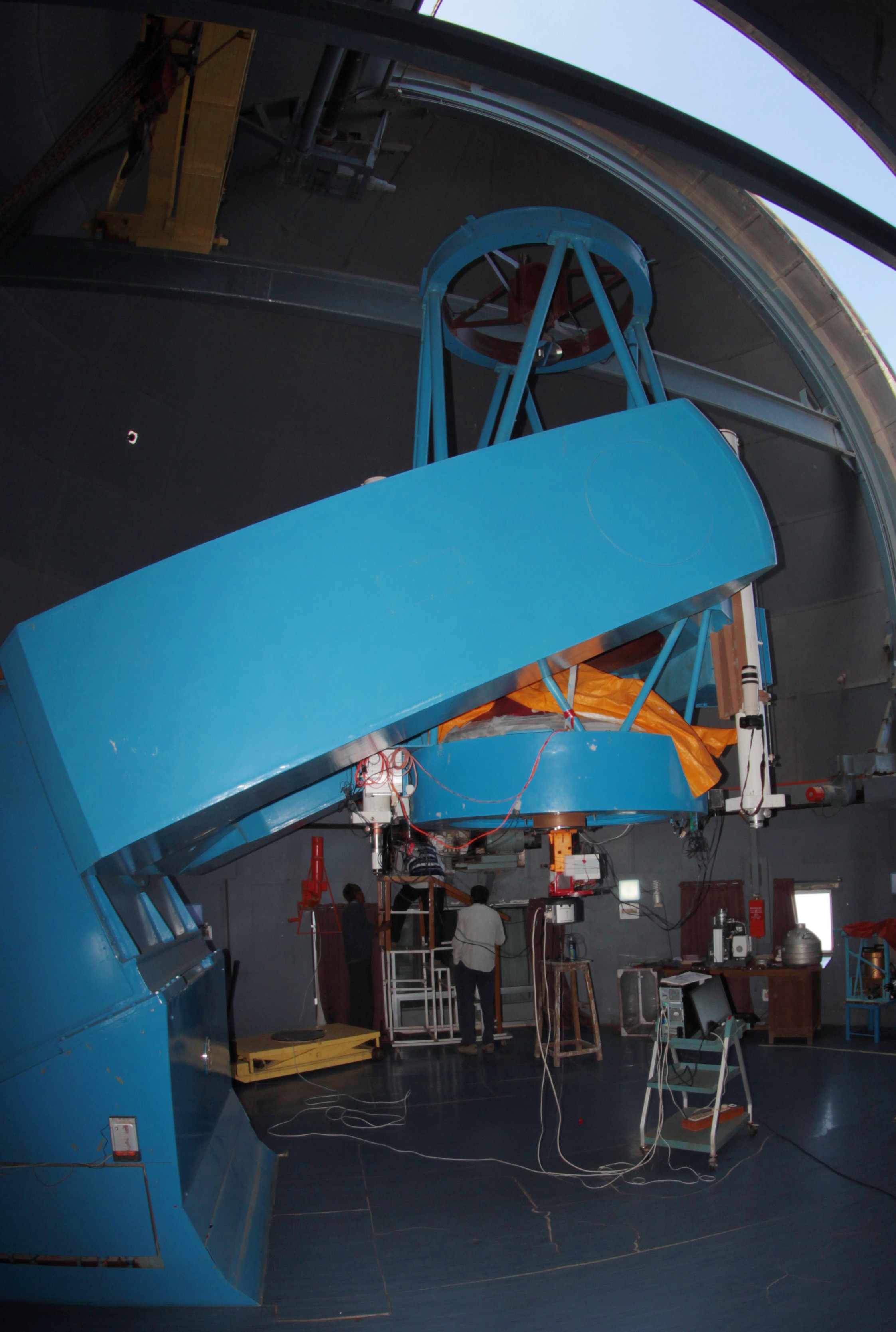}
\includegraphics[width=0.59\textwidth]{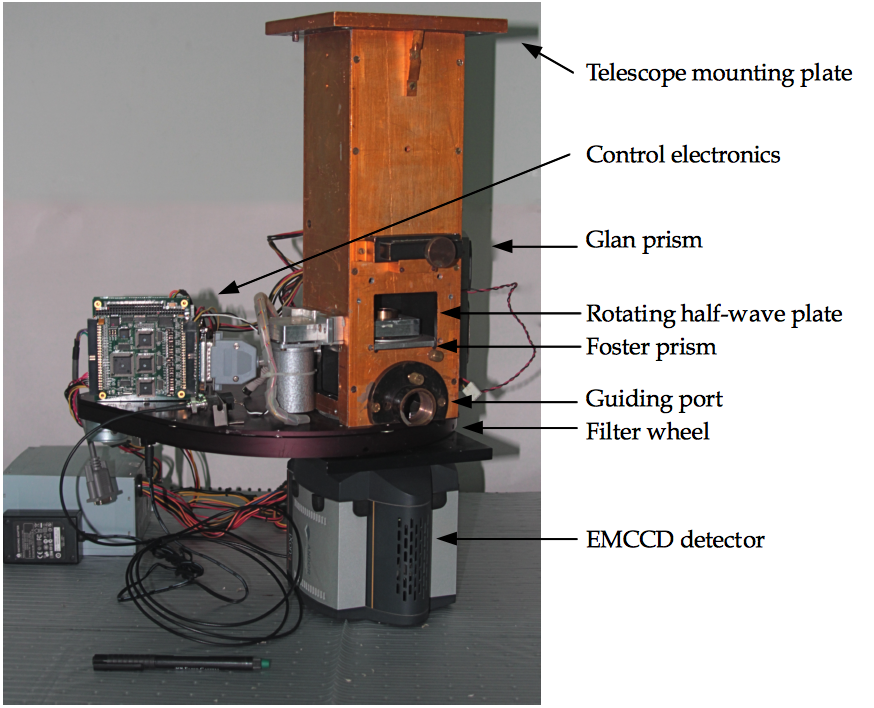}
   \end{center}
   \caption 
%>>>> use \label inside caption to get Fig. number with \ref{}
   { \label{fig:empolontel} 
Left: PRL 1.2 m MIRO telescope with the imaging polarimeter mounted at the cassegrain focus.  Right:  Annotated photo of EMPOL instrument on the optical bench during development in the lab}
\end{figure}

\noindent
A Glan prism is used to generate a 100\% linearly polarised beam for calibration purpose.  All the 3 polarisation optics are kept in the f/13 converging beam of the telescope just before the filter and EMCCD focal plane array detector.  Fig. \ref{fig:empolontel} (left panel) shows the EMPOL instrument mounted at the Cassegrain focus of the 1.2 m telescope.  The different parts of the instrument are shown in the annotated right panel of the same figure.

\subsection{Modulation and acquisition trigger electronics}

\begin{figure} [ht]
   \begin{center}
\includegraphics[width=0.38\textwidth]{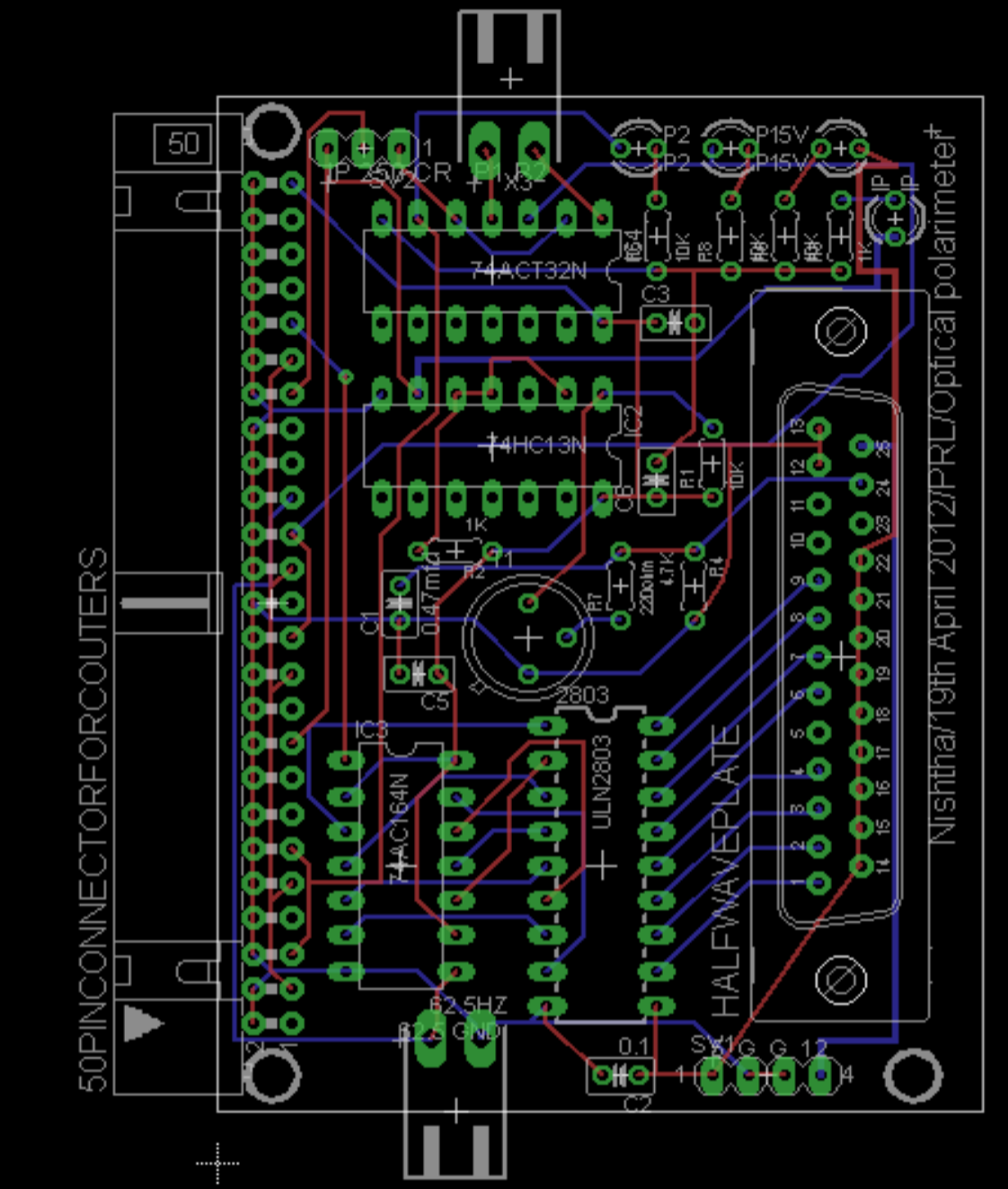}
\hfill
\includegraphics[width=0.49\textwidth]{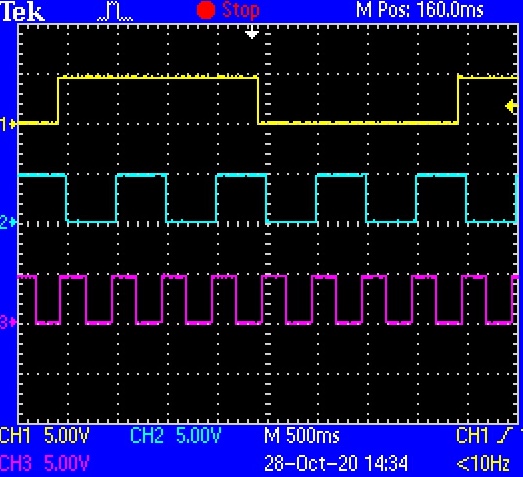}
   \end{center}
   \caption 
%>>>> use \label inside caption to get Fig. number with \ref{}
   { \label{fig:pcblayout} 
Left: PCB layout for the interface board used in EMPOL.
Right: Oscilloscope screenshot showing square waves of the three frequencies required to drive the 8-phase stepper motor.}
\end{figure}

The first version of the electronics used in this setup was shared with the optical photopolarimeter described\cite{SG2009PRLTN} earlier.  A PC104 CPU board had been used for driving the stepper motor with the help of an additional 8254 timer/counter based PC104 board to generate the three square waves of different frequencies required for rotating the stepper motor and triggering the image acquisition.  However, in a recent upgrade the entire PC104 stack was replaced with an Arduino board generating the square waves (shown in the right panel of Fig. \ref{fig:pcblayout}).  We still retain the interface board (PCB layout in the left panel of Fig. \ref{fig:pcblayout}) to drive the stepper motor as well as to trigger the image acquisition.  A 74LS164 serial-to-parallel shift register is used to generate the 8 square waves shifted by one clock pulse each.  The clock pulse has a frequency 8 times the stepper pulse frequency.  These 8 square waves are input to the ULN2803A current driver IC to drive the stepper motor.

\subsection{Zero point of position angle}

In order to define a reference position angle location for all observations, the image acquisition sequence is started at a fixed orientation of the half-wave plate.  This position is sensed using an led + photo-resistor combination across the driving gear wheel fixed to the shaft of the stepper motor.  A tiny hole in the disc of the gear wheel allows the light from the led to fall on the photo-resistor once per rotation.  The resistance is monitored by a transistor amplifier and pulse shaping circuit (upper part of the interface PCB in the left panel of Fig. \ref{fig:pcblayout}). This 'index' pulse is provided as an input to the Andor SOLIS software via the auxiliary digital input available on the Andor PCI interface card which drives the EMCCD.

\subsection{EMCCD and image acquisition scheme}

EMCCD detectors have the possibility of generating several frames per second (based on the array size and on-chip binning employed).
The Andor iXon EMCCD used in EMPOL has 1K $\times$ 1K pixels of 13 $\mu$m each.  With 4 $\times$ 4 binning over the full frame, we can achieve one frame per second with seeing limited nyquist sampling of the PSF.  Images are taken in synchronisation with every step of the half-wave plate.  The rotation of the half-wave plate is achieved with an 8-phase low power stepper motor.  48 steps complete one full rotation of the half-wave plate.  The Andor EMCCD has several modes of image acquisition.  In EMPOL, we use the external trigger mode for the image acquisition where the start of each image exposure is triggered externally using a square wave derived from the clock frequency described above.  This trigger pulse is applied as input from the Arduino to the external trigger input pin available on the EMCCD camera. 

\noindent
The AndorBASIC scripting facility available in the SOLIS software is used to complete the data acquisition.  This script takes input from the user for the total effective exposure time required (per position of the half-wave plate).  It then calculates the number of frames required to achieve the effective exposure time.  The script monitors the auxiliary digital input for the 'index' pulse and then initiates the exposures once the 'index' pulse has been detected.  Exposure times are set depending on the magnitude of the source being observed.  For faint sources, the exposures are built up by stacking the images taken at the corresponding positions of the half-wave plate to obtain the required effective exposure time.  Since this is currently completed offline, the observing procedure is quite data storage intensive.  

\noindent
A CFW-12 filter wheel with 12 slots, manufactured by Finger Lakes Instruments, is used in the light path after the Foster prism and just before the EMCCD detector.  This allows to use standard UBVRI or Sloan filters for the polarimetric measurements.  We also have the Hale-Bopp comet set of narrow-band filters\cite{Farnham2000Icarus} for use with this instrument.

\section{Observations of astronomical targets}

\subsection{Standard star observations}

\begin{figure} [ht]
   \begin{center}
\includegraphics[width=0.49\textwidth]{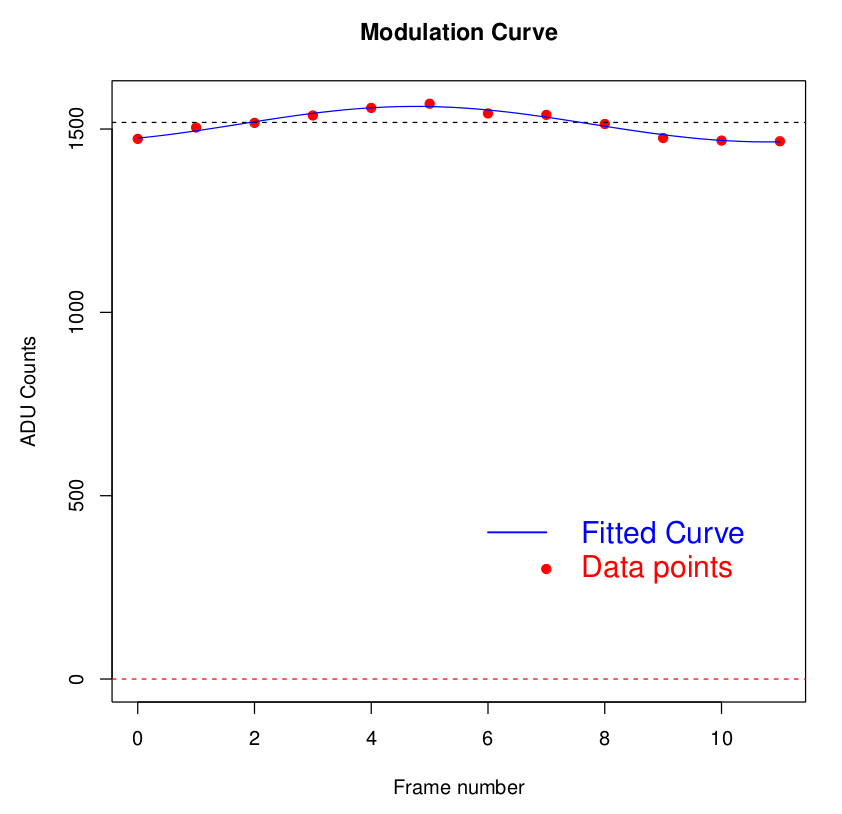}
\includegraphics[width=0.49\textwidth]{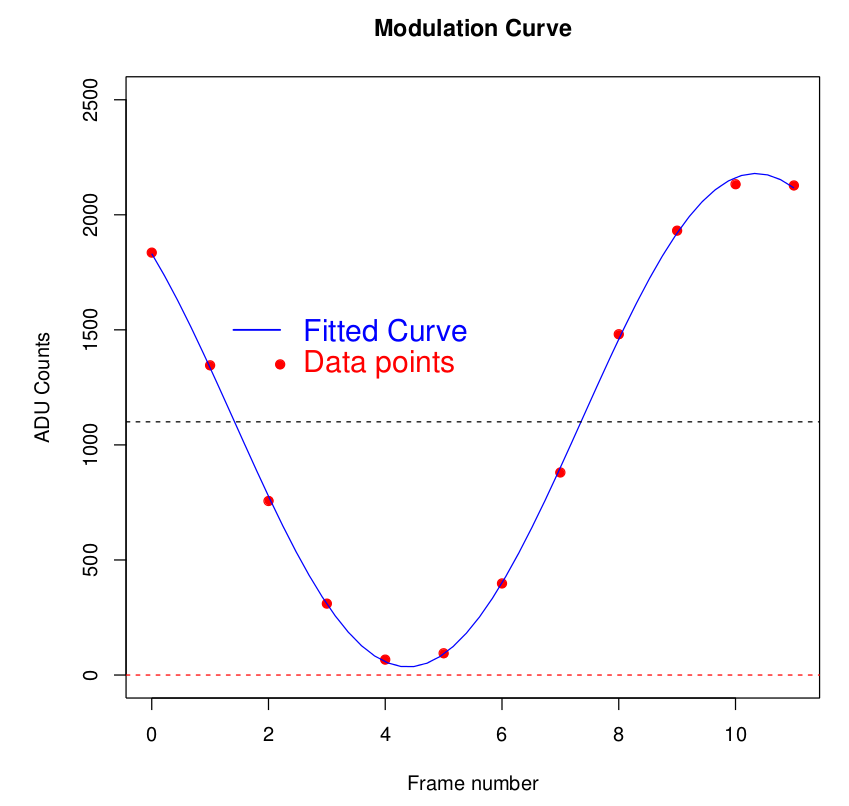}   \end{center}
   \caption 
%>>>> use \label inside caption to get Fig. number with \ref{}
   { \label{fig:9gem} 
Left: Modulation curve for polarisation standard 9 Gem.
Right:  Modulation curve for 100\% polarisation observed with Glan prism.}
\end{figure}

Polarisation standard stars are observed to calibrate the zero point of the polarisation position angle and to confirm proper operation of the instrument.  Fig. \ref{fig:9gem} (Left panel) shows the modulation curve obtained for polarisation standard 9 Gem, with 3\% polarisation.  

\noindent
To confirm that the instrument is behaving linearly over the entire range of modulation we also observe a star with a Glan prism in the light path.  The resulting polarisation due to the presence of the Glan is 100\%. The observed modulation curve is shown in the right panel of Fig. \ref{fig:9gem}.  The value derived is 97\% showing that the instrument is working in an acceptable manner.

\subsection{Blazar OJ 287}
\begin{figure} [ht]
   \begin{center}
\includegraphics[width=\textwidth]{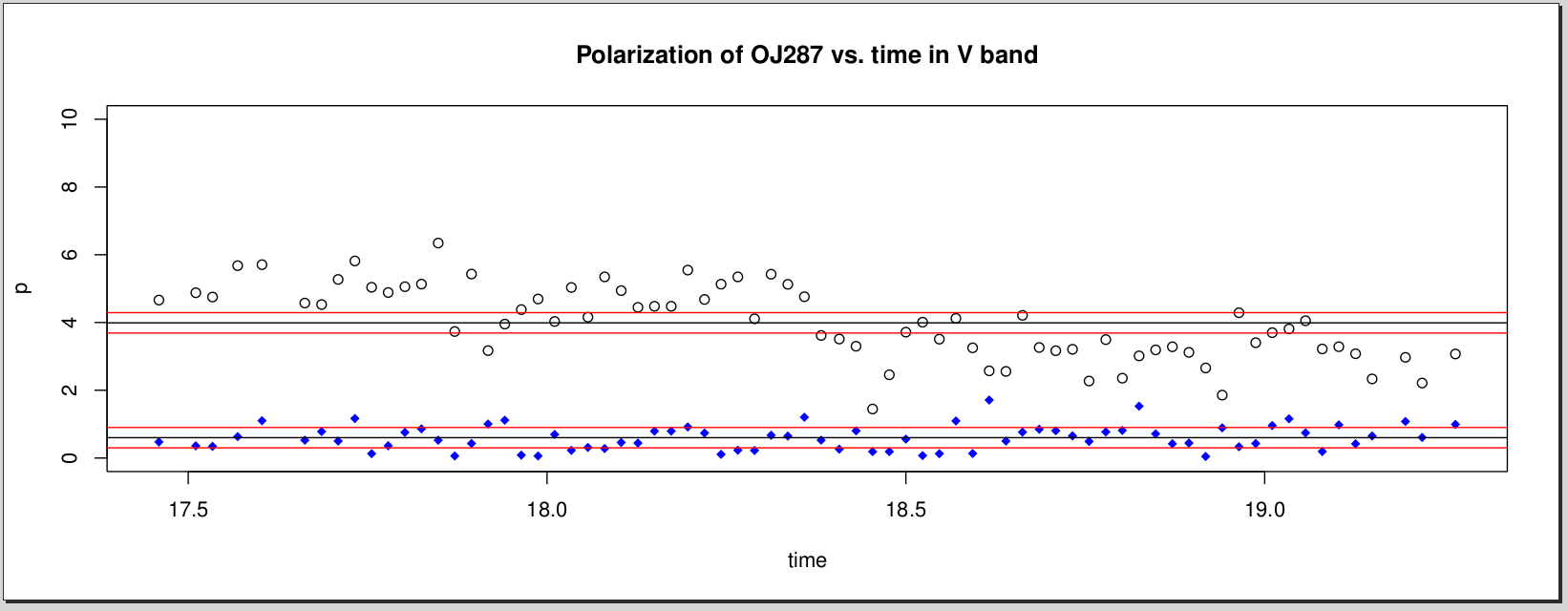}
   \end{center}
   \caption 
%>>>> use \label inside caption to get Fig. number with \ref{}
   { \label{fig:oj287} 
Polarisation variability in blazar OJ 287.  Open circles are for OJ 287 while the blue diamonds represent a comparison star nearby.}
\end{figure} 

\noindent
The well known blazar OJ 287 shows rapid variability in flux as well as polarisation within a night with very strong variations in linear polarisation. This blazar was observed with EMPOL for linear polarisation over nearly 2 hours of observations\cite{SG2014ATel}.  During the observations presented here (Fig. \ref{fig:oj287}), the  polarisation has decreased from nearly 6\% to 3\% within two hours, while the observations of a comparison star in the same field remained stable below 1\% level.  

\section{Summary}
\begin{itemize}
    \item An EMCCD based imaging polarimeter developed at PRL has been in use at the 1.2m telescope for several years with a variety of objects being observed.

    \item Linear polarisation is measured using a rotating half-wave plate as modulator and a Foster prism as analyser.

    \item Polarisation measurement is linear over the entire range as demonstrated by the observations using a Glan prism to measure 100\% polarisation.

    \item A 12 position filter wheel is used to obtain polarimetry in the standard UBVRI or Sloan filter set as well as with the Hale-Bopp comet filter set.
\end{itemize}

\section*{Acknowledgements}

We acknowledge the contribution of many summer trainees and project students in the various stages of the development of this instrument.   Work at PRL is funded by the Dept. of Space, Govt. of India.  

% References
\bibliography{report} % bibliography data in report.bib
\bibliographystyle{spiebib} % makes bibtex use spiebib.bst

\end{document}